\documentclass[11pt]{article}
\usepackage{amsmath,amssymb,amsthm}
\usepackage[margin=1.0in]{geometry}
\usepackage{amscd}
\usepackage{epsfig}
\allowdisplaybreaks
\makeindex
\begin{document}
\title{\bf {PURE DATA FOUNDATIONS OF MATHEMATICS}}
\author{%
  SAUL YOUSSEF%
  \hfil \\
  Department of Physics \\
  Boston University \\
}
\maketitle
\begin{abstract}
We propose an axiomatic foundation of mathematics based on the {\it finite sequence} as the foundational concept, rather than based 
on {\it logic} and {\it set}, as in set theory, or based on {\it type} as in dependent type theories.  Finite sequences lead to a concept of {\it pure data} which is
used to represent all mathematical objects.  As an axiomatic system, the foundation has only one axiom which
defines what constitutes a valid definition.  Using the axiom, an internal {\it true/false/undecided} valued logic and an internal language are defined, making 
logic and language-related axioms unnecessary.  Valid proof and valid computation are defined in terms of equality of pure data.  An algebra of pure data
leads to a rich theory of {\it spaces} and {\it morphisms} which play a role similar to the role of Category Theory in modern Mathematics.  As applications, 
we explore Mathematical Machine Learning, the consistency of Mathematics and address paradoxes due to G\"odel, Berry, Curry and Yablo.
\end{abstract}
%%%%%%%%%%%%%%%%%%%%%%%%%%%%%%%%%%%%%%%%%%%%%%%%%%%%%%%%%%%%%%%%%%%%%%%%

%%%%%%%%%%%%%%%%%%%%%%%%%%%%%%%%%%%%%%%%%%%%%%%%%%%%%%%%%%%%%%%%%%%%%%%%
\section{Foundation}

     Any system of reasoning must necessarily have foundational concepts which are intrinsically understood, even before a first definition.
In the case of type theories\cite{Type,Type2,HOTT,aldor}, for example, there is no definition of `type' and in Zermelo-Fraenkel Set Theory\cite{ZFC} (ZFC), there is no definition of `true', `false' or `set.'
In our case, we have one foundational undefined concept:
the {\it finite sequence}.  We assume that finite sequences and obvious statements about finite sequences are understood.  
Logic, logical values and the analogue of propositions, on the other hand, will appear as varieties of `data.'  Data is defined in terms of finite sequences as follows:
\begin{itemize}
\item {\it data} is a finite sequence of {\it codas}, and a {\it coda} is a pair of {\it data}.
\end{itemize}
If $A$ and $B$ are data, there is a natural {\it algebra of data} where $A\ B$ is the concatenation of data $A$ and data $B$ and $A:B$ is
the data formed by pairing $A$ and $B$ into a single coda.  For example, by the definition, the empty sequence of codas (written `$()$') is data.  Therefore,
the pairing of two empty sequences (written `$(:)$') is also data, and, therefore, the following sequence of three codas (:(:)(:)(:)) (:(:)) ((:):((:):)) is also data.
We refer to this as `pure data' since it is data `made from finite sequences of nothing.'
Both data in the ordinary sense (bits, bytes, language expressions) and mathematical concepts naturally appear as varieties
of pure data with simple definitions expressed in the algebra of data.  Notationally speaking, the colon operation binds from the right first and binds
less strongly than concatenation, so that $A:B:C$ means $(A:(B:C))$ and $A:B\ C$ means $(A:(B\ C))$.  We use the name `coda' both to refer 
to the pairing of two data as defined, and as the name of the axiomatic system that we are defining\cite{github}.

     In coda, answers to mathematical questions are determined by an equivalence relation `=' which is defined via a partial function from codas to data
called a `context.'  Given a context $\delta$, equality is defined by
\begin{equation}
	A\ B = \delta(A)\ B = A\ \delta(B)
\end{equation}
\begin{equation}
	A:B = \delta(A):B = A:\delta(B)
\end{equation}
for data $A$ and $B$, where the partial function $\delta$ has been extended to a function from data to data with identities.
Conceptually, $\delta$ is forced to have the same properties as the identity function on data.  Within a given context,
data $A$ is {\it empty} if $A=()$ and is {\it invariant} if $\delta(A)=A$.

     In coda, new definitions are added to a context as partial functions from codas to data.  We use a convention to guarantee that each such partial
function has it's own disjoint domain.  Let the {\it domain} of a coda $A:B$ be the data consisting of the first coda in the sequence $A$, or the empty
sequence if $A$ is empty.  A partial function mapping codas with a particular invariant domain to data is called a {\it definition}.  Definitions
can be added to a `valid' context if they do not clash, as guaranteed by the following.
\newtheorem*{remark}{THE AXIOM OF DEFINITION}
\begin{remark}  The empty context is valid.  If $\delta$ is a valid context and $d$ is a definition, and if no coda is in
the domain of both $\delta$ and $d$, then the union of $\delta$ and $d$ is a valid context.
\end{remark}
\noindent As an axiomatic system, coda is complete at this point.  This is quite a contrast from, for instance, ZFC which has ten axioms
even assuming predicate logic\cite{ZFC}.  The reason for the simplification will become clear in the sections which follow.  In Section 3, we show
that coda contains an internal logic implicit in the structure as defined.  We adopt this logic for reasoning within coda, so we do not
need axioms related to propositional or predicate logic.  In Section 4, we show that coda also contains an internal language which is introduced 
as a definition like any other.  The language has the unusual property that all byte sequences are valid syntax,
so additional language and language syntax axioms are also not necessary.  Axioms specifying valid `rules of deduction' are also not needed
because deduction, proof, and computation in coda are all determined by the above data equality.  Each of these is a merely a data sequence $A_0=A_1=A_2=\dots=A_n$,
deducing $A_n$ from $A_0$, computing $A_n$ starting with $A_0$ or proving that $A_0=A_n$, depending on one's point of view.
Unlike type systems with a Curry-Howard correspondence, proof and computation are the same thing in coda.

In the sections that follow,
`data' and `pure data' are interchangeable terms and `sequence' will always mean a finite sequence unless otherwise indicated.  Since coda
only has one axiom, we can refer to the Axiom of Definition as just `the axiom.'

\section{Atoms}

     Given fixed invariant data $D$ with length less than or equal to 1, the partial identity function 
$i_D:(D A:B)\mapsto (D A:B)$ is a definition.  By the axiom, if the domain of the current context $\delta$ is disjoint 
from the domain of $i_D$, $i_D$ can be added to the current context.  In this case, a coda $(D A:B)$ is called an {\it atom} 
since no current or future definition can remove $(D A:B)$ from a data sequence.  
Data with at least one atom in it's sequence is called {\it atomic} data, and data which is equal to $()$ is called {\it empty} data.
Equations (1) and (2) imply that empty data can not be equal to atomic data and the axiom implies that this fact remains
true independent of future definitions.  We shall use the words `always' and `never' to refer to properties that are independent of
future definition, so atomic data is `always' atomic, empty data is `always' empty and atomic data and empty data are `never' equal.

    As we shall see in Section 3, the classification of empty and atomic data is the basis for the internal logic of coda.
Atoms are also the mechanism for defining permanent data such as bits, bytes and byte strings.
Starting with an empty context, a first definition is suggested by the pure data examples above where pure data
appears to be `made of $()$ and $(:)$.'
It is clear that if we were to define $(:)\mapsto()$ all pure data would collapse to $()$.  This suggests that $(:)$ should be
mapped to $(:)$ instead.  Since $(:)$ has invariant domain $()$ we can do that by defining (:$B$)$\mapsto$(:$B$) for any coda with
domain $()$.  This makes $(:)$ both an atom and invariant, making it available as a domain for new definitions.
Continuing in this fashion, we define an invariant `0-bit' to be ((:):), and an invariant `1-bit' to be ((:):(:)).
Atoms with a 0-bit domain are conventionally bit sequences and atoms with 1-bit domains are conventionally byte sequences.
This makes ordinary byte sequences available as invariant names for new definitions.  For example, the definition `pass' maps
all domains $({\rm pass}\ A:B)$ to $B$.  Depending on the situation,
it is sometimes convenient to think of `pass' as a `command,' thinking of $B$ as
`input' and $A$ as `argument'.  Alternatively, `pass' can be thought of as a binary operator taking data $A$ and data $B$ and producing data (pass $A$:$B$).
Typical examples of definitions are shown in Tables 1 and 2.
\begin{table}
\begin{tabular}{ | l | l | l | }
Domain & Description & Action \\
\hline
pass & Pass input unchanged & (pass $A$:$B$)$\mapsto$ $B$ \\
\hline
null & Empty data for any input & (null $A$:$B$)$\mapsto$ () \\
\hline
rev & Reverse the order of data & (rev : $B$)$\mapsto$ (), if $B$ is empty \\
 &  & (rev : b)$\mapsto$ b, if b is an atom \\
 & & (rev : $B$ $C$)$\mapsto$ (rev:$B$) (rev:$A$) \\
 \hline
 if & Conditional $B$ & (if $A$:$B$) $\mapsto$ $B$, if $B$ is empty \\
 & & (if $A$:$B$) $\mapsto$ (), if $B$ is atomic. \\
 \hline
 ap & Apply argument $A$ to each input & (ap $A$ : $B$)$\mapsto$ (), if $B$ is empty \\
  & & (ap $A$ : b)$\mapsto$ $A$:b, if b is an atom \\
  & & (ap $A$ : $B$\ $C$)$\mapsto$ (ap $A$:$B$) (ap $A$:$C$) \\
\hline
nat & The natural numbers & (nat : $n$)$\mapsto$ $n$ (nat\ :\ $n+1$) \\
\hline
= & Equality & $(=A:B)\mapsto$ (), if $A=B$, atomic if $A\neq B$ \\
\hline
def & Make a new definition & (def $A$:$B$)$\mapsto$ add definition ($A$ $A'$:$B'$)$\mapsto$ $B$. \\
\hline
\end{tabular}
\caption{\label{ }{\it Typical definitions in coda.  Each definition is a partial function from codas to data defined on
codas with a specified domain.  When multiple actions are listed, the total action is defined by the first action
where the left hand side pattern applies.}}
\end{table}
In addition to the approximately 50 simple pre-defined definitions, coda includes a definition for (def $A$:$B$) which lets one
add new definitions to a context provided that $A$ is invariant and not already used in the context, according to the Axiom of Definition.
An important and typical family of definitions are the {\it applications} such as `ap' in Table 2.  These
are simple combinatorial operations on sequences.  Since finite sequences are foundationally understood, we presume that descriptions
such as those in Table 2 are sufficient.  Further description and built in examples are available with the software\cite{github}.
\begin{table}
\begin{tabular}{| l  l  l | }
Schematic Coda & $\mapsto$ & Schematic Result \\
\hline
ap $A$:$b_1$ $b_2$ $b_3$ & $\mapsto$ & ($A$:$b_1$) ($A$:$b_2$) ($A$:$b_3$) \\
app $a_1$ $a_2$ $a_3$:$B$ &  $\mapsto$ & ($a_1$:$B$) ($a_2$:$B$) ($a_3$:$B$) \\
ap2 $a$ $a_1$ $a_2$ $a_3$:$B$ & $\mapsto$ & ($a$ $a_1$:$b_1$) ($a$ $a_2$:$b_2$) ($a$ $a_3$:$b_3$) \\
aps $A$:$b_1$ $b_2$ $b_3$ $b_4$ $b_5$ $b_6$ & $\mapsto$ & ($A$ $b_1$:($A$ $b_2$:($A$ $b_3$:$b_4$($A$ $b_5$:$b_6$))) \\
apif $A$:$b_1$ $b_2$\dots & $\mapsto$ & $b_1$ (\dots if ($A$:$b_1$) is empty) $b_2$ (\dots if ($A$:$b_2$) is empty) \dots \\
\hline
\end{tabular}
\caption{\label{ }{\it Built in definitions are typically combinatoric operations on data as finite sequences.  The `ap' series are variations on the
idea of applying the `argument' A to `input' B, done in various ways.}}
\end{table}

\section{Logic}

    In the framework of coda, objects with mathematical meaning are represented by pure data.  This means that mathematical
questions will generally appear in the form
\begin{itemize}
\item Is data $A$ equal to data $B$?
\end{itemize}
Since data equality itself is available as a definition, the answers to such questions should be 
deducible from the concrete data $(=A:B)$.
This suggests that `logic' in coda should be a suitable coarse classification of data in general.  Of course, one would 
expect that a logical classification of some particular data should not change when new definitions are added via the axiom. 
The disjoint categories of empty and atomic data already meet this criteria, which suggests the definition:  
\begin{itemize}
\item data is {\it true} if it is empty, {\it false} if it is atomic and {\it undecided} otherwise.
\end{itemize}
Table 3 shows examples of this classification, including familiar binary operations from classical logic.
For instance, if data $A$ and $B$ are both either true or false, then $({\rm xor}\ A:B)$ is $()$ for {\it true} and $(:)$ for {\it false} according to the standard truth
 table for the exclusive or operation in classical logic.  On the other hand, if either $A$ or $B$ are
undecided, then the data $({\rm xor}\ A:B)$ has no definitions that apply.  In effect, $({\rm xor}\ A:B)$ `waits until both $A$ and $B$ become defined
enough' to have logical values.  A definition `bool' is available so that (bool:$A$) is () if $A$ is true, $(:)$ if $A$ is false and 
undecided otherwise. 
Although true data is always true and false data is always false, undecided data may become true or false with added definitions.
This suggests that we are defining a two valued logic where undecided data are, in a sense, variables.  This is 
not quite the complete picture because some undecided data must remain undecided independent of any added definitions.
Such {\it undecidable} data turns out to be useful in multiple ways as discussed in Sections 7 and 8. 
\begin{table}
\begin{center}
\begin{tabular}{ | c | c | }
\hline
 {\it True} & (), ({\bf pass}:), {\bf null} : a b c, ({\bf and}:), ({\bf or} a:), ({\bf xor} : a)  \\
 \hline
 {\it False} & a b c, {\bf first} 3 : a b (foo:bar), ({\bf and} a:b), ({\bf or} a:b)  \\
 \hline
 {\it Undecided} & foo:bar, {\bf pass}:foo:bar, {\bf last}:a b (foo:bar)  \\
 \hline
\end{tabular}
\end{center}
\caption{\label{ }{\it Examples of true/false/undecided data in a context where defined domains are shown with bold text.}}
\end{table}

There may be an understandable disorientation when re-defining something
as fundamental as Classical Logic.  We have found that this disorientation is only temporary once one realizes that the familiar binary logical operations remain,
 and the seemingly odd class of `undecided' data fits in perfectly as `logic valued variables' and as `answers to questions when there are no answers.'  Confidence may
 be gained in Sections 7 and 8 where we examine G\"{o}del phenomena and test the proposed logic against well known paradoxes.

\section{Language}

Both predicate logic with ZFC and dependent type theories are formal languages.  They have alphabets with special symbols
and axiomatic syntax rules which distinguish meaningful sentences from nonsense sentences such as ${\bf xx\forall yy\exists\exists}$.
The situation with coda is simpler in the sense that the language appears just as one more definition like any other.
The basic idea is to give textual control over the operations $A\ B$ and $A:B$ from the algebra of data.
Thus, if $x$ and $y$ are language expressions, partial functions
\begin{equation} \label{eqn}
( \{x\ \ y\} \ A : B ) \mapsto (\{x\}\ A : B)\ (\{y\}\ A : B)
\end{equation}
\begin{equation} \label{eqn}
( \{x : y\} \ A : B ) \mapsto (\{x\}\ A : B):(\{y\}\ A : B)
\end{equation}
allow specification of the two operations.  Language literals `A' and `B'
\begin{equation}\label{eqn}
(\{A\} A:B)\mapsto A
\end{equation}
\begin{equation}\label{eqn}
(\{B\} A:B)\mapsto B
\end{equation}
allow language specification of the left and right components of a coda.  There is more to the language, but not much more.  Additional partial functions
define grouping operations with parenthesis, allowing string literals with angle brackets, removing extraneous spaces and adding a bit of syntactic
sugar so that $A=B$ is interpreted as $(=A:B)$, $A*B:X$ is interpreted as $A:B:X$ and $X?$ is interpreted as $(?,X)$, making $X?$ behave like
a `variable.'  All of these partial functions are fused into a single partial function acting on all codas starting with text in curly braces.  The order of
the fusing determines the precedence of the operations.  Figure 1 shows an example of the language in use in a coda notebook\cite{github}.
\begin{figure}[h]
\centering
\includegraphics[width=0.6\textwidth]{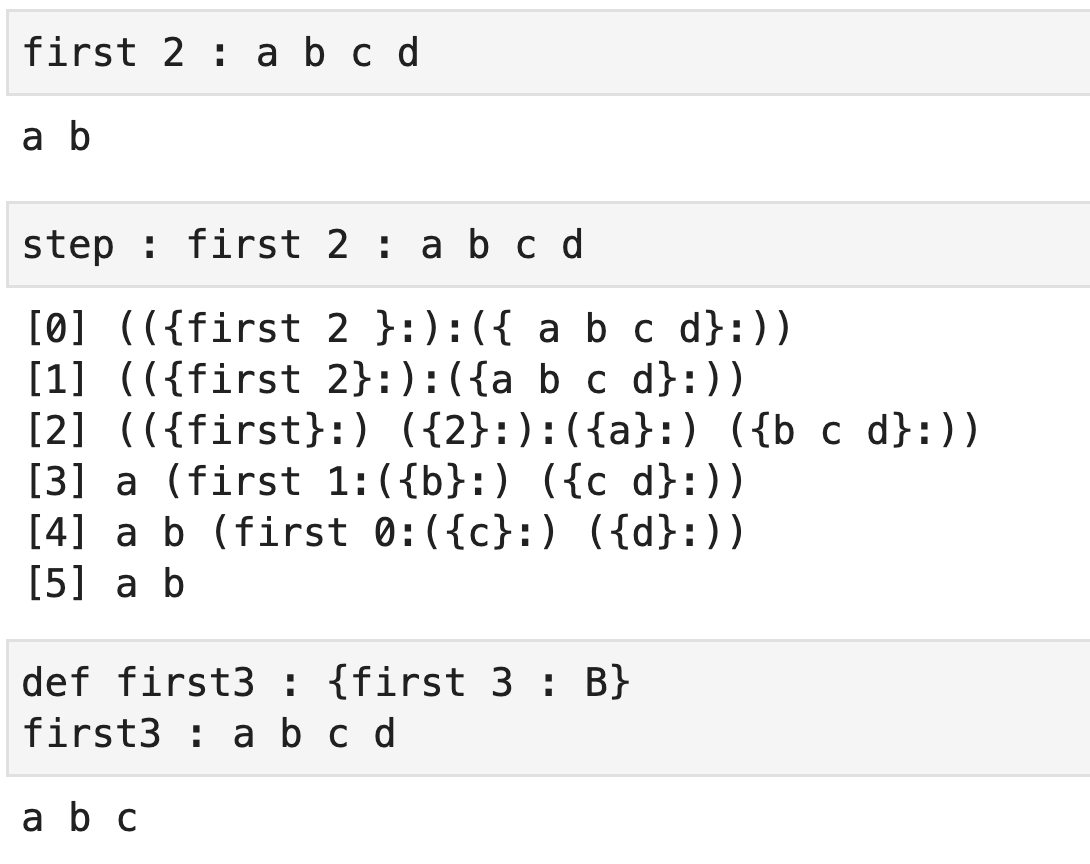}
\caption{{\it Excerpt from a coda notebook session.  The top cell shows the `first' command selecting the first two items from the `input'
sequence of four codas.  The middle cell shows the same operation using `step' to  show the individual steps evaluating the
data `\{first 2 : a b c d\}:' which is the coda definition of the language expression `first 2 : a b c d'.  Since the language is a definition
in context like any other, language operations mix with other definitions as evaluation proceeds.  The third cell shows a typical definition added within the language.}}
\end{figure}

     The language is an intentionally minimalist wrapping of the algebra of data.
Only the characters `{\bf():\{\}=*/\ }' have language significance.  There are no reserved keywords or
special syntax for variables, functions, classes or exceptions.  Any finite byte sequence $s$ can be `compiled' into pure data via $s \mapsto$ ($\{s\}$:).  
The language is unusual in that any finite byte sequence is a valid language expression.
This means that we do not have to specify an alphabet or even define valid and invalid syntax.  There is no such thing
as invalid syntax.  The source code for the compiler and parser is small enough to be easily read and understood by
humans \cite{github}.

\section{Proof and Computation}

     As noted in Section 2, coda is both a formal system and a computational system.  As a formal system, the `sentences' of the formal system
are the pure data, and the valid `rules of deduction' are that a context $\delta$ may be applied to any data according to equations (1) and (2).
Given some starting data $A_0$, for instance, any chosen coda $c$ within $A_0$ can be replaced by $\delta(c)$.  This can be repeated
as desired resulting in some sequence $A_0,A_1,\dots,A_n$.  The sequence depends on the choices, but (1) and (2) guarantee 
$A_0=A_1=\dots=A_n$ for any choice.  In this framework, the difference between a proof and a computation is merely in the strategy of where and
when to apply $\delta$.  The process of applying $\delta$ as described is called {\it evaluation}.  Most of the results in this paper
are computed with the very simple evaluation strategy of applying $\delta$ everywhere possible and then repeating until the sequence 
$A_0,A_1,\dots$ repeats or reaches a maximum chosen length. 
Figures 1 and 2 show examples of this strategy.  In Figure 1, the sequence converges to an atomic answer.  This does
not always happen, of course, as shown in the evaluation of the natural numbers in Figure 2.  Note what happens in Figure 2 when we compute
`the last natural number' (last:nat:0).  A similar result would come from computing the first of the reverse of the natural numbers (first:rev:nat:0) or the sum of all
the natural numbers (sum n:nat:0).  All these cases result in undecided data which is also undecidable.  Undecidable data is serving the
essential role of being the `answer to questions when there is no answer.'

\begin{figure}[h]
\centering
\includegraphics[width=0.45\textwidth]{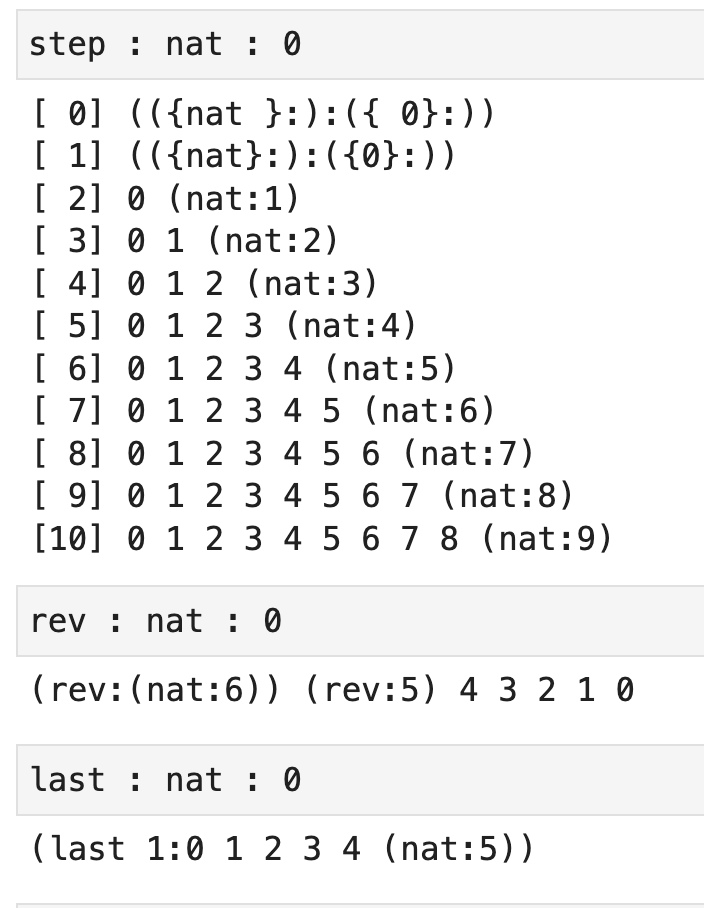}
\caption{{\it An excerpt from a coda notebook session\cite{github}.  Each cell is evaluated with the simple strategy explained in the text for up to ten
steps.  The top cell uses `step :' so each step in the strategy is displayed, showing how (nat:0) is lazily representing the natural numbers.  Note the language
elements mixing with other data in the early steps. The lower two cells show that if one computes something that has no answer, like the last natural number (last:nat:0), coda reacts by providing undecidable data as an answer.}}
\end{figure}

    The results obtained here mostly come from the above simple evaluation strategy.
We expect that there is much to be gained from a more sophisticated approach 
where one optimizes where and when $\delta$ is applied and where the equally valid $\delta(A)\mapsto A$ moves are included as well
as $A\mapsto\delta(A)$.  This is mostly unexplored territory as of this writing.  Coda has a couple of unusual and attractive computational properties.
Since $\delta$ is a partial function, it cannot `loop', by definition.  This means that an evaluation sequence $A_0,A_1,\dots$ may continue forever,
but each step in the sequence is guaranteed to terminate.  

\section{Spaces}

     Since the algebra of data is a structure common to all pure data, we might expect that the algebra plays a role analogous to the 
role that Category Theory plays in modern Mathematics\cite{Mazur}.
 Here we take a few steps towards developing this idea, aiming first at the most natural
 way of specifying collections of data of interest called `spaces' and their associated `morphisms.'  First, a few
 preliminary definitions.
 \begin{itemize}
 \item data $A$ is {\it idempotent} if $A:A:X=A:X$ for all $X$.
 \item data $A$ is {\it distributive} if $A:X\ Y=(A:X)\ (A:Y)$ for all $X$, $Y$.
 \item data $A$ is {\it abelian} if $A:X\ Y=A:Y\ X$ for all $X$, $Y$.
 \end{itemize}
 With the algebra of data in mind, any fixed data $A$ defines a collection
\begin{equation}\label{eqn}
A:X
\end{equation}
where $X$ can be any data.  We say that $(A:X)$ {\it belongs to} $A$.  Naturally,
if $(A:X)$ and $(A:Y)$ belong to $A$, we want $(A:X)\ (A:Y)$ to also belong to $A$.  This is guaranteed if
we require
\begin{equation}\label{eqn}
A : (A : X)\ ( A: Y) = A : X\ Y
\end{equation}
for all data $X$ and $Y$.  Thus, we have the definition of a {\it space}.  Given two spaces $A$ and $B$, if
distributive data $F$ satisfies
\begin{equation}\label{eqn}
F : A : X = B : F : X,
\end{equation}
then $F$ is a mapping from $A$ to $B$ where the distributivity of $F$ guarantees that $F$ respects sequences.
In this situation, we say that $F$ is a {\it morphism} from space $A$ to space $B$.  In the case where $F$
is a morphism from space $A$ to $A$, we just say that $F$ is a `morphism of $A$.'  With abuse of pure data
equality, we say that $F$ is a morphism from $A$ to $B$ if $F*A=B*F$ where $A*B:X$ is defined to be $A:B:X$.
Each space has a special element $(A:)$ which we call the {\it neutral data} or the {\it neutral element} of a space.
If $A$ is a space and if space $B$ neutralizes each $A:X$ in the sense that
\begin{equation}\label{eqn}
A : (A:X)\ (B:X) = A : (B:X)\ (A:X) =  (A:)
\end{equation}
then we call $B$ an {\it anti-space} of $A$.  A distributive space is called a {\it type}.  
Data $F$ where $F:X$ is a space is called a {\it functor}.

Table 4 shows examples of spaces and their corresponding morphisms.
\begin{table}
\begin{tabular}{| l | l | l | l | l |  }
Space & Action & Data &  $F*Space=Space*F$ \\
\hline
pass & A$\mapsto$A & All data & F distributive \\
null & A$\mapsto$ () & () only & F such that F:X=() \\
first & a A$\mapsto$ a & Single atoms &  F(a) for atom a \\
bool & A$\mapsto$() $or$ (:) & () or (:) &  F preserving logic \\
type n & A$\mapsto$ (n:3) (n:5) & Natural numbers &  (type n)$*$F$*$(type n) for any F \\
sum n & (n:5) (n:3)$\mapsto$ (n:8) & Natural numbers &  $F(n_1+n_2)=F(n_1)+F(n_2)$ \\
prod n & (n:5) (n:3)$\mapsto$ (n:15) & Natural numbers &  $F(n_1*n_2)=F(n_1)*F(n_2)$ \\
sort n & (n:5) (n:3)$\mapsto$ (n:3) (n:5) & Natural numbers & $n_1\le n_2 \implies F(n_1)\le F(n_2)$ \\
set n & (n:5) (n:3)$\mapsto$ equiv. classes & equiv. classes & F preserving classes \\
Sum n & $F_1\ F_2\dots F_n\mapsto F_1*\dots *F_n$ & Linear maps $n\mapsto n$ & Functorial \\
Sum & $F_1\ F_2\dots F_n\mapsto F_1*\dots *F_n$ & Linear maps  & Functorial \\
Space & $A\mapsto S_1 S_2\dots S_n$ & Spaces & ${\rm Space}*F*{\rm Space}$ for any $F$\\
Mor & $F_1\ F_2\dots F_n\mapsto F_1*\dots *F_n$ & Morphisms & Functorial \\
\hline
\end{tabular}
\caption{{\it Examples of spaces and their morphisms.  Of the spaces listed, only pass, null, type n and Space are types.
The neutral elements are (n:0) for sum n, (n:1) for prod n, the empty equivalence class for set n and pass for Sum n, Sum and Mor.  All others have neutral element ().}}
\end{table}
It is helpful to think of a space as doing two things:  1) selecting or constructing data of interest and 2) defining how
a finite sequence of selected data are combined to make new data in the space.
The situation with (type n) is typical.  The space (type n) constructs natural numbers from it's `input' if possible,
discarding anything that can't be converted to natural numbers represented as codas like (n:5).  The space (type n) is
distributive and, therefore, idempotent, so natural numbers in it's input are preserved.
Since (type n) is distributive, the morphisms of (type n) are just sequence preserving arbitrary functions from naturals to naturals.
The data (sum n) is also a space, selecting the same sequences of  natural numbers as (type n), but combining
sequences with addition so that sum n : (n:3) (n:5) is equal to (n:8).  Note that a space determines it's morphisms.
A morphism of (sum n) must satisfy $({\rm sum\ n}):F:X = F:({\rm sum\ n}):X$ meaning that $F$ must be linear.
Morphisms of (sort n), on the other hand, must satisfy $({\rm sort\ n}):F:X=F:({\rm sort\ n}):X$ meaning that
morphisms of (sort n) must be order preserving functions.

     It seems promising to create an abstract theory of spaces with different additional properties.  For
 example,
\begin{itemize}
\item A space with an anti-space is a {\it pure group}.
\end{itemize}
The name is justified because the set $\{G:X | $X$\ is\ pure\ data\}$
is a group in the ordinary sense where composition, defined to be $(G:X)\times(G:Y)\rightarrow G:(G:X)\ (G:Y)$, is associative, making 
($G$:) the identity, and where ($G^{-1}$:$X$) is the inverse of ($G$:$X$), assuming that $G^{-1}$ is the promised anti-space of $G$.

     An interesting possibility with spaces might be called `mathematical machine learning.'  A simple
example is illustrated in Figure 3 where we imagine that we have become interested in two different
collections of mathematical objects each just consisting of finite sequences of the atom (:).
\begin{figure}[h]
\centering
\includegraphics[width=1.0\textwidth]{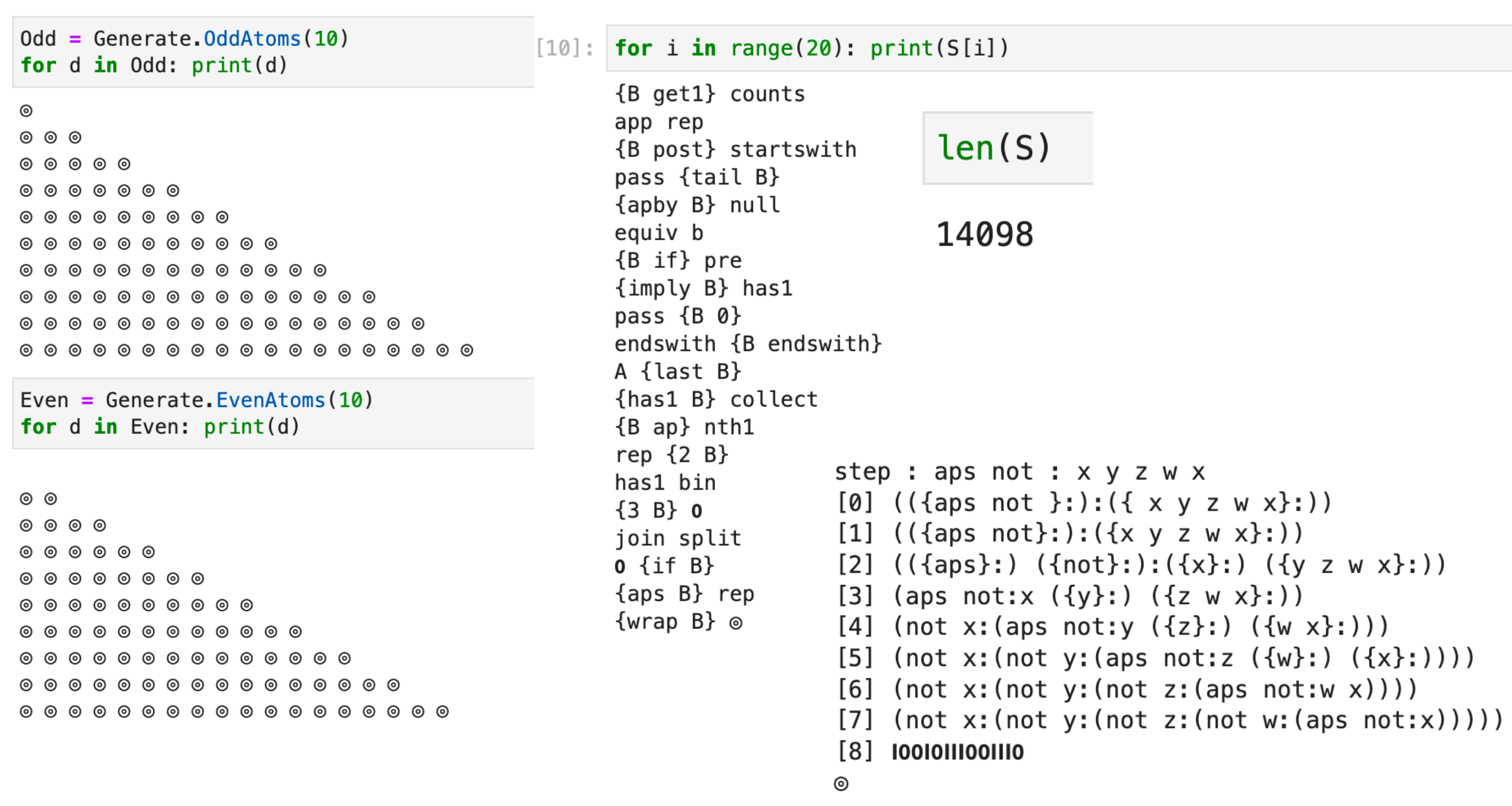}
\caption{{\it Excerpt from a coda notebook session showing a first demonstration of a situation where one is 
attempting to classify some mathematical objects of interest.  In this case, searching for a data that distinguishes a sample of
data with odd numbers of (:) atoms (displayed as circles with a small circle inside) from a sample of data with even numbers of (:) atoms.  
A search of 14,098 arbitrarily generated data (some shown), yields three successes including `aps not' as discussed in the text.}}
\end{figure}
The difference between the two collections is that one has an odd number of atoms in each sequence and
the other has an even number.  We proceed by searching for data $A$ that succeeds in classifying
the two samples in the sense that $A:X$ have different logical values for the two collections.   A first search attempt
finds three solutions out of 14,098 data, including the data  `aps not' which turns out to be true for
even numbers of (:) and false for odd numbers of (:).
Some features of the solution are interesting.
\begin{itemize}
\item The solution is clever.  It combines a combinatorial operator `aps' with a logical operator `not'
in a way that the author did not think of before doing the search.
\item The solution generalizes, since `aps not' selects data with an even number of atoms in all cases,
not just when applied to sequences of (:) atoms.
\item Although `aps not' is not quite a space, the minor modification `bool$*$(aps not)' is a space
in the above sense.
\item The morphisms of the space `bool$*$(aps not)' are interesting.  They are the functions
which preserve atomic parity of data.
\end{itemize}
Thus, just by picking out some mathematical items of interest, we have found both a way to
select similar items and an associated new mathematical structure.
Although this is merely a first toy example, this appears to be a strategy of general interest.

    Although the theory of spaces and morphisms clearly has something of the flavor of category theory, there are
also drastic differences.  In category theory, morphisms can be composed only if their domains and co-domains
agree.  In coda, {\it any} data $F$, $G$ and $H$ can be composed as 
$X\mapsto H:G:F:X$.  The products $F_1*F_2*\dots *F_n$ in Table 3,
for example, are always defined.  Another major difference
is the relationship between spaces and morphisms.  In the theory of spaces, the morphisms between space
$A$ and space $B$ are determined by $A$ and $B$.  In categories, on the other hand, there is always a choice of morphisms.  
For instance, nothing prevents one from defining a category
of real vector spaces with arbitrary functions as morphisms.  

\section{Is Mathematics Consistent?}

     We have defined empty data to be {\it true} and atomic data to be {\it false}.  But since pure data
cannot be both true and false, we have seemingly proved that coda is consistent, seemingly
in contradiction with G\"{o}del's second incompleteness theorem.  To explore this issue,
express the consistency of coda directly in coda.  The following data
\begin{equation}\label{eqn}
{\rm ap\ \{xor\ (coda:B) : (not:coda:B) \} : allByteSequences :}
\end{equation}
logically expresses the consistency of coda in the coda language.  Here the `coda' operation maps
finite byte sequences to data via the onto mapping $s\mapsto$ ($\{s\}$:).  This means that $({\rm coda}:s)$ and
$({\rm not}:{\rm coda}:s)$ will be evaluated for each byte sequence $s$ and if they ever have the same logical
value, the overall value of (11) is false.   Although each application of `${\rm xor}\ ({\rm coda}:B) : ({\rm not}:{\rm coda}:B)$'
is true in (11), we cannot quite conclude that (11) is true because (allByteSequences:) will
eventually produce the entire byte sequence (11), which will then recursively get re-evaluated
without limit.  Thus, we have the situation where (11) is equal to
\begin{equation}\label{eqn}
() () () () () \dots ()\  (some\ undecided\ data)
\end{equation}
with endless empty sequences concatenated with forever undecided data.
We know that the undecided data will never produce an atom, but we also know that it will never disappear,
meaning that (11) never quite becomes equal to (), which is the definition of `truth.'  This appears
to be a satisfying answer.  A G\"{o}del-like limitation appears, but, it is also clear that coda
is essentially consistent in the sense that no actual contradiction can appear, even though we
cannot prove that the expression `there will be no contradiction' is true in the coda internal sense.

\section{Paradoxes}

     Here we examine logical paradoxes as a way of stress testing the proposed conception of logic.
From the coda point of view, any proposition, paradox or not, should be represented as pure data.
Any pure data can be evaluated in a given context, and coda must, therefore, give a coda-logical result for 
any such proposition.

\subsection{G\"{o}del}

Recall the overall structure of G\"{o}del's second incompleteness theorem\cite{Godel}.  G\"{o}del cleverly constructs a `G\"{o}del sentence' G
in Peano Arithmetic (PA) that says of itself that G is not provable in PA.  Assuming that PA is consistent in the sense
that sentences proven in PA are actually, platonically true, we proceed to reason outside of PA in platonic logic as follows.
\begin{enumerate}
\item Suppose that G is false;
\item Then G is provable in PA;
\item Since PA is consistent, by assumption, G is platonically true;
\item Therefore G is not provable in PA.  $\Rightarrow\!\Leftarrow$;
\item Therefore, G is true.
\end{enumerate}
Thus, we conclude that G is platonically true, but G is not provable within PA.

In coda, data $X$ is provable if and only if $X=()$, so we only have to evaluate a `G\"{o}del coda' $G?$
with definition $G?\mapsto {\rm not}:G?$.   This definition is established by
\begin{itemize}
\item let G : not : G?
\end{itemize}
in the coda language.
Evaluating $G?$ a few times, we conclude that $G?$ is equal to the undecided data
\begin{equation}\label{eqn}
{\rm (not:(not:(not:(not:(not:(not:(not:(not:(not:(({not}:):({G?}:)))))))))))}
\end{equation}
which is neither true (empty) nor false (atomic).
Since further evaluation of (13) just produces more of the same, $G?$ is an example of data which is {\it undecidable}.   At this point,
G\"{o}del's argument could be repeated to conclude that $G?$ is platonically true, but unprovable in coda.  This seems less
appealing than in G\"odel's argument, just because the self-referential source of the problem is so much more obvious.  An alternative
is to adopt coda logic platonically and just consider $G?$ to be neither true nor false.

\subsection{Berry}

     One possible attitude about, say, the Liar Paradox, is to just exclude 
the paradox from mathematical consideration.  After all, not all English sentences make sense mathematically, and perhaps the liar paradox is just another such sentence.  
This point of view becomes much less convincing with Berry's paradox\cite{Berry} since Berry's proposal 
 seems like a legitimate mathematical question.  What is 
 \begin{itemize}
 \item {\it The smallest positive integer not definable in less than twelve words?}
 \end{itemize}
The problem is this.  Surely, for any positive integer $N$, $N$ either is or isn't definable in less than twelve words.
There is, therefore, a smallest integer that isn't definable in less than twelve words.  But if such an integer exists, we have succeeded in defining it in
less than twelve words.  This is a contradiction.

    Let's agree that a positive integer is `definable in coda' if it appears in the sequence (\{$s$\}:) for some 
byte string $s$.  Let (bytes:N) produce all byte sequences with length less than or equal to N and let 
(posint:B) filter for any positive integers which happen to be in it's input.  Let (berry:B) be the smallest 
positive integer not in B.  Then  
\begin{equation}
{\rm berry:posint:coda:bytes:26}
\end{equation}
computes the smallest positive integer not definable in coda in less than 26 bytes.  
Since, however, (14) is 26 bytes long, the entire string `berry:posint:coda:bytes:26' 
will eventually re-appear as an input to `coda' guaranteeing that (14) can never evaluate to a positive integer.  
Thus, (14) is in the coda logical category {\it undecidable}.  

\subsection{Curry}

Curry's paradox\cite{Curry} is another self-referential paradox.
\begin{equation}
{\rm If\ (15)\ is\ true,\ then\ Germany\ borders\ China.}
\end{equation}
We reason as follows.  Suppose (15) is true.  Then Germany borders China.  We have thus
proved `If (15) is true, then Germany border China'.  This is exactly (15).

In coda, we can define
\begin{itemize}
\item let Curry's\_sentence : imply Curry's\_sentence? : Germany\_borders\_China?
\end{itemize}
Here (imply $A$:$B$) has the standard truth table for implication when $A$ and $B$ are true and false and
is undecided otherwise.  When evaluated in coda, `Curry's\_sentence?', as expected, is forever undecided data.  It is undecidable.

\subsection{Yablo}

     Since the Liar, Berry and Curry paradoxes have self-referential sentences, it is tempting to think
that self-referential sentences is the source of the problem.  That this is not true was shown by Yablo\cite{Yablo} who
pointed out a simple paradox with no self-referential sentences.  Consider an infinite sequence
of sentences $Y_1,Y_2,\dots$ 
\begin{enumerate}
\item $Y_i$ is false for all $i>1$.
\item $Y_i$ is false for all $i>2$.
\item $Y_i$ is false for all $i>3$.
\item \dots
\end{enumerate}
Suppose that $Y_n$ is true for some $n$.  Then all sentences greater than $Y_n$ are false.  But this means
that all sentences greater than $Y_{n+1}$ are also false, which means that $Y_{n+1}$ is true, contradicting
the assumption that $Y_n$ is true.  Thus, $Y_n$ cannot be true for any $n$.  But if all $Y_n$ are all false, then
$Y_0$ is true.  This is a contradiction.

     As in the previous cases, we can proceed concretely.  Introduce a definition for Yablo as follows.
\begin{itemize}
\item def Yablo : \{ap not : Yablo : skip 1 : nat : B\}
\end{itemize}
so that $({\rm Yablo}:n)$ is true if $({\rm Yablo}:n+1), ({\rm Yablo}:n+2), \dots$ are all false.
\begin{figure}[h]
\centering
\includegraphics[width=0.9\textwidth]{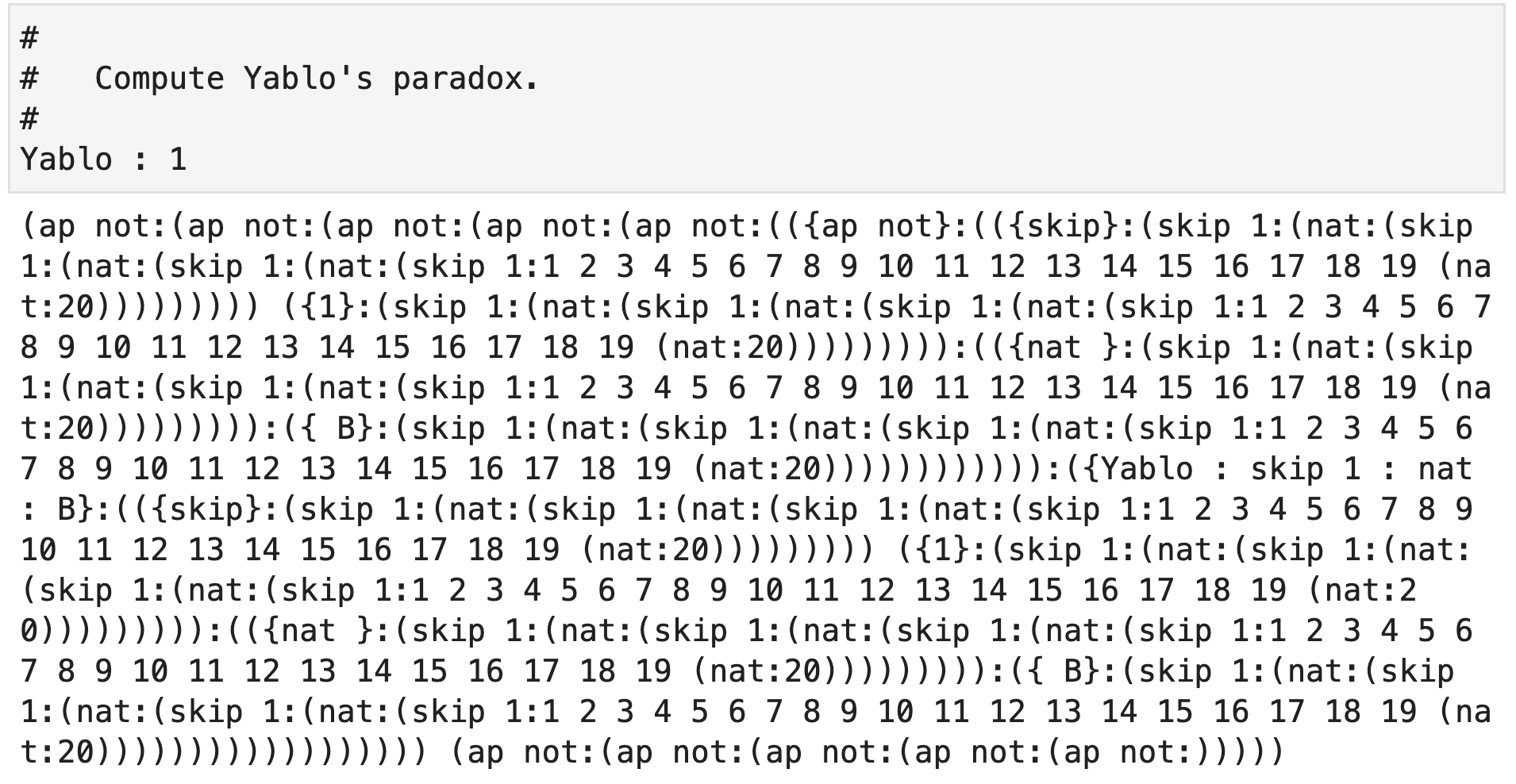}
\caption{{\it Evaluation of Yablo's proposition (Yablo:1) in a coda notebook, showing that (Yablo:1) remains undecided to depth 20, 
as might be expected.}}
\end{figure}
Figure 4 shows the standard evaluation of (Yablo:1) in coda to depth 20.  In the G\"odel case, it is `obvious' from
the output that evaluating to greater depth will change nothing.  We have not attempted to actually
prove it, but (Yablo:1) empirically remains undecided to depth 20 and we can expect that nothing changes 
at greater depth. 
\subsection {Questions that have no answers} 
    It is certainly true that questions in Mathematics may seem syntactically reasonable, but 
may actually have no answer.  For example, `Is $x=y$?' when $x$ and $y$ have not been defined, or 
`What is the last natural number?' or `What is the sum of all the natural numbers?' or 
any of the paradoxes examined in this section.  In coda, all of these are in the 
logical category {\it undecided} and all but the first are also in the category {\it undecidable}.  
Undecided data are `questions that currently have no answer' and undecidable data are 
`questions that can never have an answer.'  In a sense, the Liar paradox has no answer 
in the same way that the natural numbers have no last element.  

\section{Summary}

An axiomatic foundation for Mathematics is proposed, taking {\it finite sequence} as the starting point, rather than starting with logic, sets or types.  
In this framework, called `coda,' all mathematical quantities are {\it pure data}.  Coda has only one axiom, the {\it Axiom of Definition}, which defines 
what constitutes a valid definition.  The axiom induces a stable classification of pure data which we interpret as a logical system with 
$empty/atomic/neither$ data interpreted as $true/false/undecided$ logical values respectively.  A language is introduced via the axiom which 
gives programmatic control over pure data operations.  All byte sequences are valid language expressions so language axioms and 
axioms defining valid and invalid syntax are not needed.  Proof and computation are defined in terms of data equality.  Unlike 
dependent type systems with a Curry-Howard correspondence, proof and computation are the same thing in coda.  A theory of {\it spaces} and 
{\it morphisms} is introduced and apparently plays a role analogous to the role of Category Theory in modern  
Mathematics.  We make a first attempt at `Mathematical Machine Learning' where spaces are deduced from a classification problem. 
As an exercise of the proposed logic, we computationally address the issue of the consistency of Mathematics and paradoxes 
due to G\"odel, Berry, Curry and Yablo.  We reach a positive conclusion about the consistency of Mathematics and reach seemingly 
satisfying conclusions about the paradoxes where each paradox results in data in the undecidable category.  

    A great deal of further work seems warranted.  Only crude evaluation strategies have been attempted so far and this is an area 
where large gains are likely.  These gains are likely important for developing tools for Mathematical exploration and proof assistance.  
The example of Mathematical Machine learning in Section 6 is, at most, a hint at what's possible.  The apparently rich theory 
of spaces seems promising and has just barely started.   

%%%%%%%%%%%%%%%%%%%%%%%%%%%%%%%%%%%%%%%%%%%%%%%%%%%%%%%%%%%%%%%%%%%%%%%%
%%% \bibliography{jpsi}

\begin{thebibliography}{10}
\bibitem{Type} Dybjer, Peter and Erik Palmgren, {\it Intuitionistic Type Theory}, The Stanford Encyclopedia of Philosophy (Spring 2023 Edition), Edward N. Zalta \& Uri Nodelman (eds.)
\bibitem{Type2} Steve Awodey and Thierry Coquand, {\it Univalent Foundations and the Large-Scale Formalization of Mathematics}, Institute for Advanced Study, 2013.
\bibitem{HOTT} {\it Homotopy Type Theory: Univalent Foundations of Mathematics}, Institute for Advanced Study, 2013.
\bibitem{aldor} S.Watt, {\it Dependent Types and Categorical Programming}, TRICS, University of Western Ontario, 2012, available as {\rm https://cs.uwaterloo.ca/}$\sim${\rm smwatt/talks/2012-trics-categorical.pdf}; pp. 265-270, in Handbook of Computer Algebra J. Grabmeier, E. Kaltofen, V. Weispfenning (editors) , Springer Verlag, Heidelberg 2003 , ISBN 3-540-65466-6; S.Youssef, {\it Prospects for Category Theory in Aldor}, 2008, {\rm https://physics.bu.edu/}$\sim${\rm youssef/aldor.pdf}.
\bibitem{ZFC} Bagaria, Joan, {\it Set Theory}, The Stanford Encyclopedia of Philosophy (Spring 2023 Edition), Edward N. Zalta \& Uri Nodelman (eds.)
\bibitem{github} An implementation of coda in Python is available at {\rm https://github.com/Saul-Youssef}.  The software includes Python API, a command line 
interface and a Jupyter notebook plug-in.  
\bibitem{egg} The coda language is a descendent of {\it egg}.  See Saul Youssef, John Brunelle, John Huth, David C. Parkes and Margo Seltzer {\it Minimal Economic Distributed Computing}, arXiv:0902.4730.  J. Brunelle et al., In GECON 2006: Proceedings of the 3rd International Workshop on Grid Economics and Business Models, Singapore, 16 May 2006, ed. H. Lee, and S. Miller. Singapore; Hackensack, NJ: World Scientific.
\bibitem{Curry} Shapiro, Lionel and Jc Beall, {\it Curry’s Paradox}, The Stanford Encyclopedia of Philosophy (Winter 2021 Edition), Edward N. Zalta (ed.)
\bibitem{Godel} Raatikainen, Panu, {\it Gödel’s Incompleteness Theorems}, The Stanford Encyclopedia of Philosophy (Spring 2022 Edition), Edward N. Zalta (ed.)
\bibitem{Yablo} S. Yablo (1985). {\it Truth and reflection}. Journal of Philosophical Logic. 14 (2): 297–348. doi:10.1007/BF00249368.
\bibitem{Berry} Nicholas Griffin (2003-06-23). {\it The Cambridge Companion to Bertrand Russell}. Cambridge University Press. p. 63. ISBN 978-0-521-63634-6.
\bibitem{Mazur} Barry Mazur, {\it When is one thing equal to some other thing?}, Harvard University, 2007,  {\rm https://people.math.harvard.edu/}$\sim${\rm mazur/preprints/when\_is\_one.pdf}. 
\end{thebibliography}

%%%%%%%%%%%%%%%%%%%%%%%%%%%%%%%%%%%%%%%%%%%%%%%%%%%%%%%%%%%%%%%%%%%%%%%%
\end{document}